\documentclass[11pt]{article}
\usepackage{fullpage} 
\usepackage{graphicx}
\usepackage{epstopdf}
\usepackage{epsf}
\usepackage{epsfig}
\usepackage{amsmath}
\usepackage{amssymb}
\usepackage{dsfont}
\usepackage{comment}
\usepackage{eucal}

\newcommand{\mynoindent}{}
\newcommand{\fullpaper}[1]{}

\newcommand{\mytt}[1]{\texttt{\hyphenchar\font45\relax #1}}

\newcommand{\ie}{\emph{i.e.},}

\newcommand{\red}{\textbf{red}}

\newcommand{\blue}{\textbf{blue}}

\newcommand{\vs}{\textit{vs}}
\newcommand{\src}{\textit{source}}
\newcommand{\nonce}{\textit{nonce}}
\newcommand{\tallies}{\textit{tallies}}
\newcommand{\CC}{\textit{CC}}

\newcommand{\Texel}{\mbox{Texel}}

\newcommand{\Consensus}{\mytt{Consensus}}
\newcommand{\Implementation}{\mytt{Implementation}}
\newcommand{\NonblockingConsensus}{\mytt{NonblockingConsensus}}
\newcommand{\ProtoConsensus}{\mytt{ProtoConsensus}}

\newcommand{\UNDECIDED}{\mytt{UNDECIDED}}
\newcommand{\DECIDED}{\mytt{DECIDED}}
\newcommand{\DANGEROUS}{\mytt{DANGEROUS}}
\newcommand{\OPEN}{\mytt{OPEN}}
\newcommand{\PREDECIDED}{\mytt{PREDECIDED}}

\newcommand{\EXPERIMENTING}{\mytt{EXPERIMENTING}}

\newcommand{\SUPPORTING}{\mytt{SUPPORTING}}

\newcommand{\query}{\mytt{query}}
\newcommand{\response}{\mytt{response}}
\newcommand{\decide}{\mytt{decide}}

\newcommand{\experiment}{\mytt{experiment}}

\newcommand{\processQuery}{\mytt{processQuery}}
\newcommand{\processResponse}{\mytt{processResponse}}

\newcommand{\predecide}{\mytt{predecide}}
\newcommand{\revert}{\mytt{revert}}

\newcommand{\deliver}{\mytt{deliver}}
\newcommand{\local}{\mytt{local}}

\newcommand{\specification}{\textbf{specification}}
\newcommand{\var}{\textbf{var}}
\newcommand{\initially}{\textbf{initially}}
\newcommand{\transition}{\textbf{transition}}
\newcommand{\precondition}{\textbf{precondition}}
\newcommand{\action}{\textbf{action}}

\newcommand{\myqed}{\hfill{\rule{2.5mm}{2.5mm}}}

\newcommand{\ang}[1]{\left\langle {#1} \right\rangle}

\newtheorem{theorem}{Theorem}
\newtheorem{lemma}[theorem]{Lemma}
\newtheorem{corollary}[theorem]{Corollary}

\newenvironment{proof}[1][Proof]{\begin{trivlist}\setlength{\itemsep}{-0.3em}
\item[\hskip \labelsep {\bfseries #1}]}{\end{trivlist}}
\newenvironment{definition}[1]{\begin{trivlist}\setlength{\itemsep}{-0.3em}
\item[\hskip \labelsep {\bfseries #1:}]}{\end{trivlist}}

\newcommand{\Proof}[1]{
	\begin{proof} #1 \myqed\end{proof}
}

\title{Asynchronous Consensus Without Rounds\footnote{This paper was written in 2010.  Citations may be outdated.}}
\author{Robbert van Renesse}
\date{Department of Computer Science, Cornell University}
\begin{document}

\maketitle

\begin{abstract}
Fault tolerant consensus protocols usually involve ordered
rounds of voting between a collection of processes.
In this paper, we derive a general specification of fault
tolerant asynchronous consensus protocols
and present a class of consensus protocols
that refine this specification without using rounds.
Crash-tolerant protocols in this class use $3f+1$ processes, while
Byzantine-tolerant protocols use $5f+1$ processes.
\end{abstract}


\section{Introduction}

To the best of our knowledge,
all published crash and Byzantine
fault tolerant asynchronous consensus protocols use \emph{rounds}
(well-known examples include~\cite{Ben83,BT83,DLS88,L98,CL99},
new ones appearing frequently).\footnote{
So-called \emph{permissionless blockchains} do not use round numbers as such either, but are probabilistic and assume synchrony.}
In each round (\emph{aka} ballot, phase, or view),
a collection of processes $\cal P$ (sometimes called
\emph{acceptors}) vote on a set of proposals or abstain.
The number of rounds is unbounded.
If each process in a quorum votes for proposal $c$ in the same round,
then $c$ is decided.\footnote{Consistent with Lamport's terminology,
we distinguish the protocol \emph{deciding} from clients \emph{learning}
the decision.}
By requiring that processes vote at most once in a round and that
quorums overlap, it is impossible that different
proposals are decided in the same round.
In order to prevent different rounds from deciding different values,
rounds are totally ordered and protocols maintain the following invariant:

\begin{definition}{Round Vote Safety}
A process may vote for $c'$ in round $r'$ only if no $c$, $c \ne c'$,
is decided in a round $r$ ordered before $r'$.
\end{definition}

Based on this invariant, various generic consensus frameworks such
as~\cite{MR99,MRR02,GR04,Zie06,SvRSD08,MHS09,GKQV10,RMS10}
that claim to capture many or even all
asynchronous consensus protocols have been published.
One is tempted to conclude that fault tolerant consensus
protocols require ordered rounds.  In this paper, we show that such
a conclusion would be wrong by describing {\ProtoConsensus},
a class of asynchronous consensus protocols that does not use
rounds.
In our protocols processes also vote, and can change their votes,
but in the absence of rounds (and the state used to implement
round ordering) they cannot leverage Round Vote Safety.
Also, in the absence of rounds,
our protocols use a different rule for determining when they decide,
based on consistent cuts.

In Section~\ref{sec:spec} we describe a simple specification and model
for asynchronous consensus implementations.
We refine this specification in Section~\ref{sec:nonblocking} for
protocols that have to tolerate failures.
In Section~\ref{sec:proto} we further refine this specification
to obtain {\ProtoConsensus}, and we present
``{\Texel},'' a binary consensus protocol in this class of protocols.
Section~\ref{sec:correctness} proves the protocol's correctness.
We discuss differences with round-based protocols, learning, termination,
and coping with Byzantine failures in Section~\ref{sec:discuss}.

\section{Consensus Specification and Model}\label{sec:spec}

\begin{figure}
\begin{center}
\begin{tabular}{cc}
\begin{tabular}{c}
\fbox{\footnotesize\begin{minipage}{3in}
\begin{tabbing}
XX\=XX\=XX\=XX\=XX\=\kill
{\specification} ${\Consensus}(C)$: \\
\>{\var} \textit{decision} \\[0.5em]

\>{\transition} ${\decide}(c)$: \\
\>\>	{\precondition}: \\
\>\>\>		$c \in C \wedge \textit{decision} = {\UNDECIDED}$ \\
\>\>	{\action}: \\
\>\>\>		$\textit{decision} := {\DECIDED}(c)$
\end{tabbing}
\end{minipage}} \\[4em]
(a) \\[0.3em]
\begin{minipage}{2.5in}
\caption{
\label{fig:consensus}
(a) Specification of state transitions in {\Consensus}.
(Initially, $\textit{decision}$ is either $\UNDECIDED$ or
$\DECIDED(c)$ for some $c \in C$);
(b) Generic implementation consisting of a set of
processes $\cal P$, their local states,
a message buffer (initially $\emptyset$), and
transitions.
}
\end{minipage}
\end{tabular} & \begin{tabular}{c}
\fbox{\footnotesize\begin{minipage}{3in}
\begin{tabbing}
XX\=XX\=XX\=XXXXXX\=XXXXXXXX\=\kill
{\specification} ${\Implementation}({\cal P})$: \\
\>{\var} $\{ \, \textit{localState}_\pi \,|\, \pi \in {\cal P} \,\}, \textit{msgbuffer}$ \\[0.5em]

\>{\initially}: $\textit{msgbuffer} = \emptyset$ \\[0.5em]

\>{\transition} ${\deliver}(\pi, m, \textit{args})$: \\
\>\>	{\precondition}: \\
\>\>\>		$\ang{\pi, m} \in \textit{msgbuffer} ~\wedge$ \\
\>\>\>		$\texttt{someCondition}(\textit{localState}_\pi, m, \textit{args})$ \\
\>\>	{\action}: \\
\>\>\>		$\textit{msgbuffer} := \textit{msgbuffer} ~\backslash~ \{ \ang{\pi, m} \}$ \\
\>\>\>		$(\textit{localState}_\pi, \textit{msgs}) :=
					\texttt{someFunc}(\textit{localState}_\pi, m, \textit{args})$ \\
\>\>\>		$\textit{msgbuffer} := \textit{msgbuffer} \cup \textit{msgs}$ \\[0.5em]

\>{\transition} ${\local}(\pi, \textit{args})$: \\
\>\>	{\precondition}: \\
\>\>\>		$\texttt{someCondition}'(\textit{localState}_\pi, \textit{args})$ \\
\>\>	{\action}: \\
\>\>\>		$(\textit{localState}_\pi, \textit{msgs}) :=
					\texttt{someFunc}'(\textit{localState}_\pi, \textit{args})$ \\
\>\>\>		$\textit{msgbuffer} := \textit{msgbuffer} \cup \textit{msgs}$
\end{tabbing}
\end{minipage}} \\
\\
(b)
\end{tabular}
\end{tabular}
\end{center}
\end{figure}

Figure~\ref{fig:consensus}(a)
shows a simple state transition specification for a consensus protocol
(excluding interface transitions for proposing and learning).
A specification defines states and gives legal state transitions
between those states.
The specification as a whole, as well as individual transitions, can involve
parameters (listed in parentheses) that are bound within the defined scope.
A {\transition} definition includes a {\precondition} and an {\action}.
If the precondition holds then the transition is \emph{enabled}.
A transition is performed indivisibly, including validating its precondition
and executing the action should the precondition hold.
Transitions are deterministic,
but multiple transitions may be enabled
simultaneously.
An \emph{execution from state $s_0$} complies with a specification if it
involves a series of transitions, where the first transition is enabled
in $s_0$, and each subsequent transition is enabled in the state produced
by performing the previous transition.
If an execution from $s_0$ exists that results in a state $s$, we say
that $s$ is \emph{reachable} from $s_0$.

In the specification ${\Consensus}(C)$ of Figure~\ref{fig:consensus}(a),
$C$ is a set of \emph{candidate values}.
For simplicity of exposition, $C$ is fixed and $|C| > 1$.
Variable \textit{decision}
completely characterizes the state of {\Consensus}, and
is either {\UNDECIDED} or ${\DECIDED}(c)$ for some $c \in C$.
There is a transition ${\decide}(c)$ for each candidate value $c$.
The transition is enabled only if $c$ is an element of $C$
and
if $\textit{decision} = {\UNDECIDED}$ holds.
The ${\decide}(c)$ transition causes {\Consensus} to move from
state ${\UNDECIDED}$ into state ${\DECIDED}(c)$.
It is easy to derive the following property:

\begin{definition}{Consistency}
{\Consensus} cannot transition into both ${\DECIDED}(c)$
and ${\DECIDED}(c')$ if $c \ne c'$.
\end{definition}

\noindent
But we also require the following
to disqualify implementations that never decide,
as well as implementations that can only decide the same predetermined
element in $C$:

\begin{definition}{Non-triviality}
For each $c \in C$, state ${\DECIDED}(c)$ is reachable
from state {\UNDECIDED}.
\end{definition}

This paper focuses on implementations of ${\Consensus}(C)$ that employ a
collection of processes $\cal P$ communicating over a network.
These implementations can be specified in much the same way, as illustrated in
Figure~\ref{fig:consensus}(b).
The \emph{implementation state} is a collection of local states, one
for each process, and a message buffer, which is
a set of (\textit{destination process identifier}, \textit{payload}) pairs.
The first argument of each transition is a process identifier.
Transitions at a process $\pi$ can read and write the process's local
state $\textit{localState}_\pi$, but not that of others.
Transitions are also allowed
to add $\ang{\textit{process identifier}, \textit{payload}}$ pairs
to the message buffer.
Certain transitions, such as $\deliver$, can remove such a pair
and operate on the local state of the corresponding process.
Others, such as $\local$, do not take a message as input.
It is easy to see that simultaneously enabled transitions at different
processes are independent of one another:

\begin{lemma}\label{lemma:diamond}
If, in an implementation state $s$,
a transition $t_1$ is enabled at process $\pi_1$ and transition $t_2$
at process $\pi_2$, then first performing $t_1$ and then $t_2$, or
first performing $t_2$ and then $t_1$, result in the same implementation
state $s'$.  Also, performing one of $t_1$ or $t_2$ in state $s$ does not
disable the other.
\end{lemma}

\noindent
We make, initially,
the following assumptions about when such process transitions are taken:

\begin{itemize}\setlength{\itemsep}{-0.3em}
\item \textbf{Crash Failures}:
A process follows its specification until the host on which it runs
crashes and ceases to execute transitions.
Hosts are assumed to fail independently of one another.
A host that never fails is called \emph{correct}, as is a process
that runs on a correct host.

\item \textbf{Fairness}:
It cannot happen that, at some point, a transition becomes
permanently enabled but is not thereafter executed.
There is no bound on the time before an enabled transition
is executed (known as \emph{asynchrony}).
\end{itemize}

\noindent
In consensus protocols, the {\Consensus} state ({\ie} \textit{decision})
is some function of the implementation state~\cite{AL88}.
If this function maps the implementation state to ${\DECIDED}(c)$,
we say that the protocol decided $c$.
If the implementation state maps to {\UNDECIDED}, we say that the protocol
is undecided.

To tolerate host failures, an implementation has to distribute
the {\Consensus} state across multiple processes running on different hosts
in a way so that
the failure of some hosts does not make the {\Consensus} state
inaccessible to the others.
A protocol that is always guaranteed to
reach a ${\DECIDED}(c)$ state
would be desirable but unachievable in our environment~\cite{FLP85}.
We focus on protocols that
decide under favorable conditions,
even in the face of some limited number of crash
failures.

\section{Non-blocking Consensus}\label{sec:nonblocking}

\mynoindent
In this section, we derive a state transition specification of
\emph{Non-blocking} consensus protocols, a specification
that all fault-tolerant consensus protocols must refine.
We assume that the number of failures is bounded
by a constant $f$, $f > 0$,
and define a \emph{Fail-Prone System} $\cal F$ to be a set
consisting of all sets of $f$ processes:
${\cal F} = \{ F ~|~ F \subset {\cal P} \wedge |\,F\,| = f \}$.

Define a $\Pi$-execution to be an execution of a consensus protocol
that involves only transitions at processes in $\Pi$
($\Pi \subseteq {\cal P}$).
An undecided implementation state $s$ is defined to be \emph{$c$-decidable
by $\Pi$}
if there exists a $\Pi$-execution from $s$ leading to an implementation state
that is ${\DECIDED}(c)$.
If $s$ is $c$-decidable by $\Pi$, then $s$ is also $c$-decidable
by any superset of $\Pi$.
We say that a consensus protocol is
\emph{Non-blocking} under $\cal F$ iff:

\begin{definition}{Non-blocking}
For any undecided implementation state $s$ and any $F \in {\cal F}$,
there exists a candidate value $c \in C$ such
that $s$ is $c$-decidable by ${\cal P} \,\backslash\, F$.
\end{definition}

\noindent
For Consistency, all $c$-decidable implementation states have the following
property:

\begin{theorem}\label{theorem:disjoint}
If implementation state $s$ is $c$-decidable by $\Pi$, then
for every set $\Pi'$ of processes disjoint from $\Pi$, and
every $c'$, $c' \ne c$, $s$ is not $c'$-decidable by $\Pi'$.
\end{theorem}
\Proof{
By contradiction, suppose there exists a $\Pi$-execution $E$ from $s$ that
decides $c$, and a $\Pi'$-execution $E'$ from $s$ that decides $c'$, where
$c' \ne c$ and $\Pi \cap \Pi' = \emptyset$.
Because $\Pi \cap \Pi' = \emptyset$, the implementation state resulting from
first executing $E$ from $s$ and then $E'$ must be identical to the
implementation state resulting from first executing $E'$ and then $E$
(a corollary of Lemma~\ref{lemma:diamond}).
However, first executing $E$ causes the protocol to decide $c$ while
first executing $E'$ causes the protocol to decide $c'$.  Because an
execution of a consensus protocol
cannot decide multiple candidate values (Consistency),
we have derived a contradiction.
}

\begin{corollary}\label{corol:disjoint}
If a state $s$ is both $c$-decidable by $\Pi$ and $c'$-decidable by
$\Pi$, $c \ne c'$, then any execution from $s$
that decides any candidate value $c'' \in C$ involves transitions at
processes in $\Pi$.
\end{corollary}
\Proof{
Because $c \ne c'$,
it must be the case that $c'' \ne c$, or $c'' \ne c'$, or both.
Because $s$ is $c$-decidable, any execution that
decides $c''$, $c'' \ne c$, must involve processes in $\Pi$ (implied by
Theorem~\ref{theorem:disjoint}).
By the same reasoning, any execution that decides $c''$, $c'' \ne c'$
must involve processes in $\Pi$.
Therefore, any execution that
decides any $c''$ must involve processes in $\Pi$.
}

\noindent
We call an undecided implementation state \emph{dominated by $\Pi$}
if every deciding execution involves processes in $\Pi$.
If an implementation state is dominated by $\Pi$, then the crash of all
processes in $\Pi$ makes it impossible for the protocol to decide.
A Non-blocking consensus protocol cannot have states that are
dominated by sets of processes that might all fail (\emph{i.e.}, elements
of $\cal F$), and this is
the basis for a refinement of {\Consensus}.
We use $\cal F$ to partition undecided implementation states:

\begin{itemize}\setlength{\itemsep}{-0.3em}
\item {\DANGEROUS}: there exists a set of processes
$F$ in $\cal F$ such that the
implementation state is dominated by $F$.
\item ${\PREDECIDED}(c)$:
(i) the implementation state is not {\DANGEROUS}, (ii) there exists
an $F \in {\cal F}$ such that the implementation state is $c$-decidable by $F$,
and (iii) there does not exist an $F' \in {\cal F}$ and a
$c' \in C, c \ne c'$, such that the implementation state
is $c'$-decidable by $F'$.  (In other words, there exist $f$ processes
that can decide $c$ by themselves, while it would take more than $f$
processes to decide any other value.)
\item {\OPEN}: the implementation state is
neither {\DANGEROUS} nor {\PREDECIDED}(c) for any $c \in C$.
\end{itemize}

\begin{figure}
\begin{center}
\begin{tabular}{cc}
\fbox{\footnotesize\begin{minipage}{3in}
\begin{tabbing}
XX\=XX\=XX\=XX\=XX\=\kill
{\specification} ${\NonblockingConsensus}(C, {\cal F})$: \\
\>{\var} \textit{nbdecision} \\[0.3em]

\>{\transition} ${\predecide}(c)$: \\
\>\>	{\precondition}: $c \in C \wedge
	\forall c': \textit{nbdecision} \ne {\DECIDED}(c')$ \\
\>\>	{\action}: $\textit{nbdecision} := {\PREDECIDED}(c)$ \\[0.3em]

\>{\transition} ${\revert}(c)$: \\
\>\>	{\precondition}: $\textit{nbdecision} = {\PREDECIDED}(c)$ \\
\>\>	{\action}: $\textit{nbdecision} := {\OPEN}$ \\[0.3em]

\>{\transition} ${\decide}(c)$: \\
\>\>	{\precondition}: $\textit{nbdecision} = {\PREDECIDED}(c)$ \\
\>\>	{\action}: $\textit{nbdecision} := {\DECIDED}(c)$
\end{tabbing}
\end{minipage}} & \begin{minipage}{1.8in}
\caption{\label{fig:nonblocking} State transitions in Non-blocking consensus.
(Initially, \textit{nbdecision} is $\OPEN$ or, for some $c \in C$, either
$\PREDECIDED(c)$ or $\DECIDED(c)$.)
}
\end{minipage}
\end{tabular}
\end{center}
\end{figure}

\vspace{-1em}
\begin{theorem}\label{theorem:undecided}
If an undecided implementation state $s$ is $c$-decidable by $\{ \pi \}$ (a singleton set consisting of $\pi$), then $s$ is
either \emph{${\PREDECIDED}(c)$} or is dominated by $\{ \pi \}$ and
thus \emph{\DANGEROUS}.
\end{theorem}
\Proof{
Let $s$ be an undecided implementation state that is $c$-decidable by $\{ \pi \}$.
It is impossible that $s$ is a ${\PREDECIDED}(c')$ state if
$c' \ne c$ (by requirement (iii) on {\PREDECIDED} states).
By definition of fail-prone systems, $\pi$ must be in at least one
element $F$ of $\cal F$, and thus $s$ is $c$-decidable by $F$.
There are two cases.
(1) If $s$ is $c'$-decidable by $F$ for some $c' \ne c$, then $s$ is dominated
by $F$ (Corollary~\ref{corol:disjoint}) and thus a {\DANGEROUS} state.
(2) If no such $c'$ exists, then $s$ is a ${\PREDECIDED}(c)$ state.
}

Because Non-blocking consensus protocols cannot have states that
are {\DANGEROUS}, it must be in state ${\PREDECIDED}(c)$
just before such a protocol can decide a candidate value $c$
(by Theorem~\ref{theorem:undecided}).
The {\NonblockingConsensus} specification in
Figure~\ref{fig:nonblocking} is designed to meet exactly these requirements.
It is straightforward to exhibit a refinement mapping that shows that
{\NonblockingConsensus} is a refinement of
{\Consensus}.

\section{\ProtoConsensus}\label{sec:proto}

Most consensus protocols avoid {\DANGEROUS} states using rounds,
enforcing that within a round, if two process
vote for a candidate value (rather than abstain),
then they vote for the same candidate value.
Such consensus protocols enforce this by having two phases in each round,
one in which a value is proposed, and a second in which processes
vote on a proposed value.
For example, in the first phase, a designated leader may propose a value
that satisfies Round Vote Safety, and in the second phase the processes
vote for that value.
This approach avoids close ties while voting, which
could lead to blocking in the face of failures.
Such protocols
can tolerate fewer than half of processes in $\cal P$ failing.

In this section we present {\ProtoConsensus}, a class of crash-tolerant
consensus protocols that use voting without rounds and
avoid {\DANGEROUS} states by using two-thirds majorities.
When close to deciding one value, at least
one-third of processes have to change their vote in order
to decide another value.
As we shall see, the protocols can tolerate fewer than one-third of processes
in $\cal P$ failing, that is, $f = \lfloor (n - 1) / 3 \rfloor$,
where $n = |\,{\cal P}\,|$.

In these protocols, each process $\pi$ proposes a candidate value
by making an initial vote, but may change its vote based on the
outcome of \emph{experiments} it runs.
We say that $\pi$ \emph{supports} $c$ if $c$ is the candidate value that
$\pi$ most recently voted for.
An experiment consists of $\pi$ sending a {\query} message to
its peers and awaiting corresponding {\response} messages.
A process may run only one experiment at a time.
In between experiments, we say that $\pi$ is \emph{idle}; during
an experiment we say that $\pi$ is \emph{experimenting}.
A peer process $\pi'$ may only respond if it is idle;
a {\response} message contains the candidate value that $\pi'$
is supporting.
If $\pi$ supports $c$, but responses from one-third or more
of peer processes contain a value other than $c$,
then $\pi$ may terminate its experiment and vote for
any candidate value in $C$.
Alternatively, process $\pi$ may abort an experiment at any time
without changing the value that it is supporting.

As in~\cite{Lam78},
we define a strict partial order on experiments $x_1$ and $x_2$.
$x_1 \prec x_2$ if $x1 \ne x2$ and
\vspace{0em}
\begin{itemize}\setlength{\itemsep}{-0.3em}
\item $x_1$ and $x_2$ are experiments of the same
process, and $x_1$ happened before $x_2$; or
\item $x_1$ is an experiment of process $\pi_1$, and $x_2$ an experiment
of $\pi_2$, and the {\query} message of $x_1$ (sent by $\pi_1$)
was received and processed by $\pi_2$ before it initiated $x_2$; or
\item $\exists x: x_1 \prec x \wedge x \prec x_2$ (transitive closure).
\end{itemize}

\begin{figure}
\begin{center}
\begin{tabular}{cc}
\begin{tabular}{c}
\fbox{\footnotesize\begin{minipage}{3in}
\begin{tabbing}
XX\=XX\=XX\=XXXXXX\=XXXXXXXX\=\kill
{\specification} ${\Texel}(f)$: \\
\>{\var} $\{\, {\vs}_\pi, {\cal N}_\pi \,|\, \pi \in {\cal P} \,\},
								\textit{msgbuffer}$ \\[0.5em]

\>{\transition} ${\experiment}(\pi, c, \nonce, \tallies[\,])$: \\
\>\>	{\precondition}: \\
\>\>\>		${\vs}_\pi = {\SUPPORTING}(c) \wedge \nonce \in {\cal N}_\pi
				\wedge \tallies[c] = 1 \wedge \tallies[\overline{c}] = 0$ \\
\>\>	{\action}: \\
\>\>\>		${\vs}_\pi := {\EXPERIMENTING}(c, \nonce, \tallies)$ \\
\>\>\>		${\cal N}_\pi := {\cal N}_\pi \,\backslash\, \{ \nonce \}$ \\
\>\>\>		$\textit{msgbuffer} := \textit{msgbuffer} \cup
				\{ \ang{\pi', \ang{{\query}, \pi, \nonce}} ~|~
						\pi' \in {\cal P} \,\backslash\, \{ \pi \} \}$ \\[0.5em]

\>{\transition} ${\processQuery}(\ang{\pi', \ang{{\query}, \pi, \nonce}}, c)$: \\
\>\>	{\precondition}: \\
\>\>\>		$\ang{\pi, \ang{{\query}, \pi, \nonce}} \in
												\textit{msgbuffer} ~\wedge$ \\
\>\>\>\>		$({\vs}_\pi = {\SUPPORTING}(c) \vee {\vs}_\pi = {\EXPERIMENTING}(c, \cdot, \cdot))$ \\
\>\>	{\action}: \\
\>\>\>		$\textit{msgbuffer} := \textit{msgbuffer} ~\backslash~ \{ \ang{\pi', \ang{{\query}, \pi, \nonce}} \}$ \\
\>\>\>		${\vs}_\pi := {\SUPPORTING}(c)$ \\
\>\>\>		$\textit{msgbuffer} := \textit{msgbuffer} \cup \{ \ang{\pi, \ang{{\response}, \nonce, c}} \}$ \\[0.5em]

\>{\transition} ${\processResponse}(\ang{\pi, \ang{{\response}, \nonce, c'}}, c, \tallies[\,])$: \\
\>\>	{\precondition}: \\
\>\>\>		$\ang{\pi, \ang{{\response}, \nonce, c'}} \in \textit{msgbuffer} ~\wedge$ \\
\>\>\>\>		${\vs}_\pi = {\EXPERIMENTING}(c, \nonce, \tallies)$ \\
\>\>	{\action}: \\
\>\>\>		$\textit{msgbuffer} := \textit{msgbuffer} ~\backslash~ \{ \ang{\pi, \ang{{\response}, \nonce, c'}} \}$ \\
\>\>\>		\textbf{if} \>
				$\tallies[c'] = f \rightarrow {\vs}_\pi := {\SUPPORTING}(c')$ \\
\>\>\>		[] \>		$\tallies[c'] < f \rightarrow \tallies[c'] := \tallies[c'] + 1 ~~~~~ \textbf{fi}$
\end{tabbing}
\end{minipage}} \\
(a)
\end{tabular} & \begin{tabular}{c}
\includegraphics[width=1.8in]{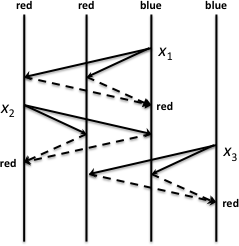} \\
(b) \\
\\
\\
\\
\begin{minipage}{1.8in}
\caption{\small \label{fig:proto} (a) Specification of {\Texel}.
(b) Event-time diagram of
three experiments $x_1$, $x_2$, and $x_3$.  Here $x_1 \prec x_2$,
but $x_1$ and $x_3$, as well as $x_2$ and $x_3$, are concurrent.
$x_1$ and $x_3$ are reversing; $x_2$ is not.  Delayed messages
are not shown.}
\end{minipage}
\end{tabular}
\end{tabular}
\end{center}
\end{figure}

\noindent
See Figure~\ref{fig:proto}(b) for examples.
The $\prec$ relation between experiments is asymmetric in
that it is impossible for two experiments $x_1$ and
$x_2$ that both $x_1 \prec x_2$ and $x_2 \prec x_1$ hold.
However, neither may hold, and we say that the experiments
are \emph{concurrent}.
We call an experiment of some process $\pi$
\emph{a reversing experiment} iff (1)
the experiment terminated; (2)
just before the experiment $\pi$ supported $c$ for some $c \in C$; and
(3) right after the experiment $\pi$ supported $c'$, with $c' \ne c$.

A \emph{consistent cut} $\CC$ of the execution of a protocol is
a collection of its reversing experiments, such that if an experiment
is a member of $\CC$, then so are all reversing experiments that precede
it according to the partial order $\prec$ defined above.
\textbf{The protocol has decided $c$ iff the execution
contains a consistent cut $\CC\,$ in which more than two-thirds of
processes support $c$ after their last experiment in $\CC$.}
We say that $\CC\,$ supports $c$ in that case.

For illustration
we describe a specific instance of {\ProtoConsensus},
a protocol that we call \emph{\Texel}.\footnote{Texel
(Lat.~53.017, Long.~4.798),
pronounced \emph{Tessel},
is a charming Dutch island, well north of
the charming Greek island of Paxos (Lat.~39.199, Long.~20.184).}
We use $C = \{ \red, \blue \}$ (that is, the protocol is binary) and
$n = 3f+1$ processes ($f > 0$).
If $c$ is an element of $C$, then by $\overline{c}$ we denote the opposing
candidate value of $c$.
In Figure~\ref{fig:proto}(a), we show the specification of {\Texel}.
Process $\pi$ maintains two variables:
\begin{itemize}\setlength{\itemsep}{-0.3em}
\item ${\cal N}_\pi$, an unbounded collection of \emph{nonces}.
These collections are disjoint, \emph{i.e.}, $\pi \ne \pi' \Rightarrow
{\cal N}_\pi \cap {\cal N}_{\pi'} = \emptyset$.
Nonces are not ordered in any way.
\item ${\vs}_\pi$: the vote state of process $\pi$,
either ${\SUPPORTING}(c)$ or
${\EXPERIMENTING}(c, \nonce, \tallies[\,])$.
Here $c$ is an element of $C$,
$\nonce$ an element in ${\cal N}_\pi$ before the experiment started, and
$\tallies[\,]$ a two-element array of non-negative integers such that
$\tallies[\red] \le f \wedge  \tallies[\blue] \le f$.
Initially, ${\vs}_\pi = \SUPPORTING(c_\pi)$, where $c_\pi$ is the proposal
of process $\pi$.
\end{itemize}

\noindent
At any time, a process $\pi$ with ${\vs}_\pi = \SUPPORTING(c)$
can set ${\vs}_\pi$ to ${\EXPERIMENTING}(c, \nonce, \tallies[\,])$,
starting a new experiment ({\transition}
${\experiment}(\pi, c, \nonce, \tallies[\,])$
in Figure~\ref{fig:proto}(a)).
Here

\begin{itemize}\setlength{\itemsep}{-0.3em}
\item $\nonce$ is an arbitrary element of ${\cal N}_\pi$,
and removed from ${\cal N}_\pi$ so it cannot be reused.
(The cardinality of ${\cal N}_\pi$ remains the same,
so no notion of ordering can be derived from its cardinality);
\item $\tallies[c] = 1$ and $\tallies[\overline{c}] = 0$.
\end{itemize}

\noindent
As part of this {\experiment} transition,
process $\pi$ broadcasts a $\ang{\query, \pi, \nonce}$ message
to its $3f$ peers (no self-delivery).
The nonce serves to recognize response messages to
this specific {\query}.

In {\ProtoConsensus} protocols, processes are not allowed to respond
to queries if they are experimenting themselves.  In {\Texel} this is
enforced by terminating an experiment upon receipt of a {\query} message
before responding.
Thus when processing a $\ang{\query, \src, \nonce}$ message
({\transition} {\processQuery} in Figure~\ref{fig:proto}(a)),
a process $\pi'$ with ${\vs}_{\pi'} = \EXPERIMENTING(c', \cdot, \cdot)$
aborts its experiment by setting ${\vs}_{\pi'} := \SUPPORTING(c')$.
If $\pi'$ already has ${\vs}_{\pi'} = \SUPPORTING(c')$, then it leaves its
state unchanged.
In either case, $\pi'$ sends $\ang{\response, \nonce, c'}$ to $\src$.

On receipt of $\ang{\response, \nonce, c'}$
({\transition} {\processResponse}), $\pi$
transitions as follows:
\begin{itemize}\setlength{\itemsep}{-0.3em}
\item if ${\vs}_\pi = {\EXPERIMENTING}(c, \nonce, \tallies[\,])$
and $\tallies[c'] = f$,
then $\pi$ sets ${\vs}_\pi := \SUPPORTING(c')$;
\item if ${\vs}_\pi = {\EXPERIMENTING}(c, \nonce, \tallies[\,])$ and
$\tallies[c'] < f$,
then process $\pi$ transitions to
${\EXPERIMENTING}(c, \nonce, \tallies'[\,])$
such that $\tallies'[c'] = \tallies[c'] + 1$, and
$\tallies'[\overline{c'}] = \tallies[\overline{c'}]$;
\item in all other cases, the process ignores the message and
remains in the same state.
\end{itemize}

\noindent
An experiment by a correct process is guaranteed to terminate---termination
requires
at most $2f$ responses, and there are $3f$ peers to provide them of which
at most $f$ are faulty.

\vspace{-0.5em}
\section{Correctness}\label{sec:correctness}

Because {\Texel} does not have rounds, it does not support
Round Vote Safety.  Instead, {\Texel} and other protocols in {\ProtoConsensus}
satisfy the following important invariant:

\begin{definition}{Consistent Cut Vote Safety}
If a consistent cut supports $c$, then no process $\pi$ that supports $c$
on that cut can subsequently run a reversing experiment.
\end{definition}
\Proof{
Suppose not, and let $\CC\,$ be a consistent cut that supports $c$
and $x$ be the first subsequent reversing experiment by a process
$\pi$ supporting $c$, setting ${\vs}_\pi$ to $\SUPPORTING(c')$, $c' \ne c$.
At the start of $x$, $\pi$ broadcasts a $\ang{\query, \pi, \nonce}$,
using a previously unused $\nonce$.
If received by some process $\pi'$, then processing must be after all
experiments of $\pi'$ in $\CC$ (if not, and processing happens before some
experiment $x'$ by $\pi'$ in $\CC$,
then $x \prec x'$ and $x$ must be part of $\CC$,
contradicting that $x$ follows $\CC$).
Before setting ${\vs}_\pi$ to $\SUPPORTING(c')$,
$\pi$ must have received
more than $f$ responses containing a value other than $c$.
Since there are only $3f$ peers of $\pi$ and at least $2f$ of its
peers must have been supporting $c$ in $\CC$, at least
one of the supporting processes in $\CC$ must have finished
a reversing experiment before $x$.
This contradicts that $x$ is the first reversing experiment.
}

\noindent
Next we show that {\Texel} supports
Consistency and Non-triviality, and is Non-blocking under $\cal F$.

\begin{theorem}\label{theorem:consistency}
\emph{\Texel} satisfies Consistency.
\end{theorem}
\Proof{
We have to prove that, in any execution, if two consistent cuts
support $c_1$ and $c_2$ resp., then $c_1 = c_2$.
Suppose not, and consider an execution in which there exists two
consistent cuts $\textit{CC}_1$ and $\textit{CC}_2$,
with $2f+1$ processes that support $c_1$ in
$\textit{CC}_1$ and $2f+1$ processes that support $c_2$ in
$\textit{CC}_2$, $c_1 \ne c_2$.
Because there are only $3f+1$ processes, there must be $f$ processes
that are supporting $c_1$ in $\textit{CC}_1$ and $c_2$
in $\textit{CC}_2$.
Since the experiments at a single process are totally ordered, each of these
processes has had a reversing experiment, contradicting
Consistent Cut Vote Safety.
}

\begin{theorem}\label{theorem:nontriviality}
\emph{\Texel} satisfies Non-triviality.
\end{theorem}
\Proof{
We have to show that
from any undecided implementation state $s$ either candidate value can be decided.
We will construct an execution from $s$ that decides (wlog) {\red}:

\begin{enumerate}\setlength{\itemsep}{-0.3em}
\item from $s$, start experiments on all idle processes so that
all processes are experimenting.
\item deliver all {\query} messages to all processes.
In {\Texel}, an experimenting process that receives a {\query} aborts
its experiment without changing its vote.
Because none of these experiments are reversing, the state is still undecided.
Thus at least $f+1$ processes must have ${\vs}_\pi = \SUPPORTING(\red)$
and $f+1$ must have ${\vs}_\pi = \SUPPORTING(blue)$
(because if only $f$ processes had ${\vs}_\pi = \SUPPORTING(c)$, then
$2f+1$ processes would have ${\vs}_\pi = \SUPPORTING(\overline{c})$
and $\overline{c}$ would be decided, contradicting that the state is undecided).
\item start experiments on all processes that have ${\vs}_\pi = \SUPPORTING(\blue)$,
deliver the corresponding {\query} messages to the processes that
have ${\vs}_\pi = \SUPPORTING(\red)$, and deliver the corresponding responses.
This causes all experiments to be reversing, and all
corresponding processes to have ${\vs}_\pi = \SUPPORTING(\red)$.
\end{enumerate}
Now that all processes have ${\vs}_\pi = \SUPPORTING(\red)$, and the
state consisting of all experiments forms a consistent cut, $\red$ is decided.
}

\begin{theorem}\label{theorem:nonblocking}
\emph{\Texel} is Non-blocking under $\cal F$.
\end{theorem}
\Proof{
Consider any undecided implementation state $s$ of the protocol and suppose $f$
processes have failed.
We show that there exists a deciding execution that does not 
involve the failed processes.
Because there are $2f+1$ correct processes and only two
candidate values, there must be
at least $f+1$ correct processes that have ${\vs}_\pi = \SUPPORTING(c)$ or
${\vs}_\pi = \EXPERIMENTING(c, \cdot, \cdot)$ for some $c \in C$.
Wlog., assume $c = {\red}$.
The deciding execution is then as follows:

\begin{enumerate}\setlength{\itemsep}{-0.3em}
\item from $s$, start experiments on all idle correct processes so that
all correct processes are experimenting.
\item deliver all {\query} messages to all correct processes.  This causes
experiments on correct processes to finish, but none will be reversing.
At this time all correct processes will be idle.
Because no experiment is reversing, the state is still undecided and
at least $f+1$ correct processes have ${\vs}_\pi = \SUPPORTING(\red)$.
\item start experiments on all correct processes that have
${\vs}_\pi = \SUPPORTING(\blue)$,
deliver the corresponding {\query} messages to the correct processes that
have ${\vs}_\pi = \SUPPORTING(\red)$, and deliver the corresponding responses.
This causes all running experiments on correct processes to be reversing,
and all those processes to have ${\vs}_\pi = \SUPPORTING(\red)$.
\end{enumerate}
Now that $2f+1$ processes have ${\vs}_\pi = \SUPPORTING(\red)$,
$\red$ is decided.
}

\noindent
{\Texel} is a refinement of {\NonblockingConsensus}.
A refinement mapping could be as follows:
\begin{quote}
The {\Texel} protocol is in a ${\DECIDED}(c)$ state if the execution
contains a consistent cut $\CC$ that supports $c$.
If no such consistent cut exists, the implementation state
is undecided.
If the implementation state is undecided and there exists a
continuation of the execution that decides $c$ with only $f$
or fewer processes making transitions, but it takes more than $f$
processes to decide any $c'$, $c' \ne c$, then the protocol is in a
${\PREDECIDED}(c)$ state.  If no such continuation exists,
the implementation state is $\OPEN$.
\end{quote}

\noindent
From Theorems~\ref{theorem:consistency}, \ref{theorem:nontriviality},
and~\ref{theorem:nonblocking} it is easy to see that this mapping is
well-defined.
Consequently, all transitions taken by the protocol either
correspond to transitions in {\NonblockingConsensus} or are
\emph{stutter transitions} (\emph{i.e.}, transitions that leave
the state of {\NonblockingConsensus} unchanged).
In particular, any process transition that leaves the implementation
state undecided
must correspond to a $\predecide$, a $\revert$, or a stutter transition,
depending on how the implementation state changes.
A process transition that causes the first existence of a consistent cut
that supports $c$ must be a $\decide(c)$ transition,
while all future process transitions are stutter transitions.

\section{Discussion}\label{sec:discuss}

\subsection{Rounds versus Consistent Cuts}

Rounds generalize synchronous \emph{communication-closed layers}~\cite{EF82}
for asynchronous environments.
The \emph{basic round model} was formalized in~\cite{DLS88}, and later
extended by papers such as~\cite{Gaf98, BS09}.
An elegant way of structuring fault tolerant asynchronous
distributed algorithms,
rounds are totally ordered and uniquely identified by round numbers.
In a round, a process
exchanges messages with other processes and takes an execution
step before proceeding to another round.
At the same point in time, different processes may be executing in
different rounds.
Each process maintains the round number of the round that it is in,
and tags messages with this number.
Messages received early (\emph{i.e.}, with a round number higher than
the round number of the receiving process) are buffered and delivered when the
process reaches that round, while messages received
late are discarded.
In~\cite{DLS88}, the authors propose a \emph{Global Stabilization Time}
after which no messages arrive late, making it possible to guarantee
termination properties.

Instead of rounds,
{\Texel} and other protocols in the {\ProtoConsensus} class use
consistent cuts defined on partially ordered experiments run
by individual processes.
Processes do not agree on rounds and corresponding round numbers.
Without a round number, {\Texel} processes have no way of detecting
and delaying delivery of early messages.
We show that indeed it is not possible to derive a round number
from the state that a {\Texel} process keeps.
To wit, a process $\pi$ maintains only its vote and
a set of nonces in between experiments.
The cardinality of this set remains unchanged throughout protocol
runs and cannot be used to derive a round number.
During an experiment, a process also uses a counter
to count responses and a nonce to match responses to the
experiment's query.
Both are forgotten after the experiment and thus
cannot be used
to derive a round number either.

A previous approach to reduce synchronization among processes in
consensus protocols was to let processes be involved in multiple
rounds simultaneously~\cite{HMRM01}.
Our approach eliminates rounds altogether, further reducing constraints
on processes.

\subsection{Learning and Termination}

A practical implementation of {\Texel} must provide a
fault-tolerant way for
processes to learn the outcome of a decision by determining
if there exists a consistent cut along which $2f+1$ processes are
supporting the same candidate value $c$.
One approach is to associate a \emph{Vector Clock}~\cite{SM94b}
with each vote.
A vector clock is a vector with an integer entry for each process
and is a compact representation of causal relationships.
Each process $\pi$ has its own copy $\textit{VC}_\pi$, initialized
to all zeroes, associated with its initial vote.
When $\pi$ starts an experiment, it increments $\textit{VC}_\pi[\pi]$,
and piggybacks the new vector clock on the experiment's {\query} message.
On receipt of $\ang{\query, \pi, \nonce, \textit{VC}_\pi\,}$
by a process $\pi'$, $\pi'$ sets
$\textit{VC}_{\pi'}$ to $\texttt{max}(\textit{VC}_{\pi'}, \textit{VC}_\pi)$
(the pairwise maximum of $\textit{VC}_{\pi'}$ and $\textit{VC}_\pi)$.
When the experiment ends, $\pi$ associates its vote with its current
vector clock.
A \emph{learner} periodically queries all processes for votes along with the
associated vector clocks.
If the learner receives votes from a set of processes $\Pi$ such that
$\forall \pi, \pi' \in \Pi: \textit{VC}_{\pi'}[\pi] \le \textit{VC}_\pi[\pi]$,
then it knows that it has collected votes from some consistent cut.
Moreover, if $|\Pi| > 2f$ and the votes are unanimously for some
candidate value $c$, then the learner learns that $c$ has been decided.

{\Texel} does not coordinate progress toward termination
(\emph{i.e.}, that learners eventually learn a decision).
For a pragmatic {\ProtoConsensus} protocol, chances of termination
need to be significantly improved, for example,
by having {\ProtoConsensus} transitions be guided by messages sent by a
weakly elected leader that attempts to have only processes
in the minority run experiments, have them query the majority, and
then vote for their candidate to forge a decision.
Such a strategy would make a decision likely, but may still fail.

To go further and guarantee termination,
the asynchronous execution model needs to be extended with
an \emph{oracle} that eventually
constrains when to perform enabled
transitions in order to forge a decision, such as the weak
ordering oracles described in~\cite{PSUC02}.
({\Texel} is reminiscent of the R-Consensus or OneThirdRule
protocol~\cite{PSUC02,BS09,RMS10}, which also uses $3f+1$ processes
but is round-based.)
Non-blockingness under $\cal F$
guarantees that terminating executions exist, and hence
such oracles can be defined, but doing so is outside the scope of this paper.

\subsection{Byzantine Failures}

The protocols we have described tolerate up to $f$ crash or
omission failures.
We briefly describe how
they can be generalized for Byzantine
failures.  To tolerate $f$ Byzantine failures, a Byzantine {\Texel}
protocol would need $5f+1$ processes communicating using authenticated links.
The protocol decides when more than $3f$ \emph{correct}
processes support a particular value
(which can only be learned if more than $4f$ processes \emph{claim} to be
supporting that value---up to $f$ processes may be lying).
A process can change its
vote $c$ if in an experiment it learns that more than $2f$ processes are
supporting a candidate value other than $c$.
This rule ensures Consistent Cut Vote Safety
because if $3f+1$ correct processes support $c$, then at most $2f$ can
claim to support a value other than $c$.
The protocol resembles Bosco~\cite{SvR08} but does not use rounds.

It is easy to see that Byzantine {\Texel} satisfies Non-triviality.
To see why the protocol is Non-blocking under $\cal F$, consider an undecided
state.
Because there are at least 4f+1 correct processes and the protocol is binary,
there are $2f+1$ correct processes supporting the same value $c$.
The remaining correct processes
could each do an experiment and end up voting for $c$ themselves,
causing $c$ to be decided (and, importantly
so, causing $c$ to be learnable).
(While the protocol is Non-blocking under $\cal F$, an adversary that
controls the order in which {\Texel} transitions happen can easily
prevent the protocol from deciding.  Again, an oracle is required to
ensure termination.)

A Byzantine-tolerant protocol that uses only $3f+1$
processes can be constructed by applying the translation method
of~\cite{HDvR07} to the crash-tolerant {\Texel} protocol.
Prior translation methods for asynchronous systems
such as described in~\cite{Bra87a} and~\cite{Coa88}
work only for round-based protocols.
While the resulting protocol still has no rounds of voting on
a candidate value, the mechanisms
of~\cite{HDvR07} does require
processes to maintain monotonically increasing counters.

\section{Conclusion}

This paper has demonstrated the existence of a class of asynchronous
consensus protocols, {\ProtoConsensus}, that is fundamentally different from
prior published consensus protocols in that it does not use rounds.
We have derived {\ProtoConsensus} through stepwise refinement from a simple
specification of consensus.
For purposes of illustration, we have described {\Texel},
a simple protocol in this class.
While {\Texel} is not a pragmatic protocol for $f > 1$ (for $f = 1$ {\Texel}
decides after at most one reversing experiment),
it represents a different breed of fault tolerant asynchronous
consensus protocols that provides new insight into the consensus problem
and merits further study.
For example, because {\ProtoConsensus} protocols do not have to agree on
round numbers, they appear candidates for self-stabilizing protocols
(protocols that would, however, not satisfy {\NonblockingConsensus}).

{
\bibliographystyle{plain}
\bibliography{all}
}

\end{document}